# Half-filled Hubbard ring with alternating site potentials in a magnetic field

*Sanjay Gupta*[†], *Shreekantha Sil*[‡] *and Bibhas Bhattacharyya*[††]

[†]TCMP Division, Saha Institute of Nuclear Physics, 1/AF-Bidhannagar, Calcutta – 700064, India.

[‡]Department of Physics, Visvabharati, Shantiniketan, Birbhum

[††]Department of Physics, Scottish Church College, 1 & 3, Urquhart Square, Calcutta – 700006

## Abstract

We have studied a Hubbard ring with alternating site potentials for half-filling in presence of a magnetic flux. Using a mean-field approach we have calculated the conductivity of such a ring at low and high temperatures. The interplay of the correlation, the polarizing field and the chemical modulation in site potentials tune the conductivity in an interesting fashion. In presence of the modulation in the site energy an appreciable variation in the conductance is observed with the change in flux. This scenario gets substantially modified in presence of the Hubbard correlation. Finite-size effects are also identified and they are found to be quickly disappearing with increasing system size. Sharp changes in the magnetoconductance is found to disappear at higher temperatures.

# I. Introduction

The role of an externally applied magnetic field on the electrical transport properties of a metal has been extensively studied for a long time [1]. Recent studies of unusual magnetoresistive properties of low-density Si MOSFET's [2] and $p$-GaAs/AlAs heterostructures [3] have generated a great deal of interest from both theoretical and experimental points of view [4]. Quasi-two-dimensional organic systems e.g. $\beta''$-(BEDT-TTF)$_2$SF$_5$CH$_2$CF$_2$SO$_3$ also show interesting field-dependence of the electrical transport properties [5]. A similar type of field-dependent phenomenon is observed very recently in highly oriented pyrolitic graphite [6]. Although a clear theoretical picture is still to emerge in this area the recent studies clearly point out that the electronic correlation is going to play a very important role in the magnetoresistivity of some of these dilute two-dimensional (2D) systems (e.g. Si MOSFET) [4].

On the other hand persistent currents in mesoscopic metallic rings have been well studied for a long time from both theoretical [7] and experimental [8] perspectives. Such a ring, penetrated by a magnetic flux, is found to show a typical Aharonov-Bohm type of oscillation [9] in the current. The presence of disorder is expected to enrich the problem [10, 11, 12] due to impurity scattering. More recently the effect of electron-electron interaction [13, 14] on mesoscopic conductance has generated a great deal of interest. Also a huge activity is going on with the manganite perovskites (e.g. La$_{1-x}$Sr$_x$ MnO$_3$ or La$_{1-x}$Ca$_x$MnO$_3$ ) for their noticeable magnetoresistive properties [15] known as colossal magnetoresistance.

In view of the scenario mentioned above we investigate the effect of the magnetic field on the conductivity of a mesoscopic ring which otherwise shows strong effects of electronic correlation in its ground state. A Hubbard chain with a periodic variation of the site potential is found to undergo a spin density wave (SDW)/ charge density wave (CDW) transition at half-filling [16, 17], and the conductivity of such a chain of finite length captures interesting features of the competition between the SDW and the CDW [17] instabilities. The effect of an externally applied magnetic flux on the conductivity in such a ring is, therefore, expected to reveal some significant aspects of mesoscopic systems.

The present problem requires some suitable approximate scheme for solution. Here we have adopted a mean field approximation (namely, the generalized Hartree-Fock approximation, GHFA in short) to tackle the problem. It is to be mentioned here that a mean field approach was found to be very much useful for capturing the SDW/CDW transition in the present model in absence of the magnetic field [16, 17].

## II. The Model and the GHFA calculations

The chemically modulated Hubbard chain has alternate sites occupied by $A$ and $B$ types of atoms. We take a finite chain of $N$ sites, where $N$ is an even number, in the form of a ring which is threaded by an externally applied magnetic field $H$; the Hamiltonian looks like:

$$\mathcal{H} = \sum_i \epsilon_i n_i + t \left( e^{i\phi_1} \sum_{i\sigma} c^\dagger_{i\sigma} c_{i+1\sigma} + h.c. \right) + U \sum_i n_{i\uparrow} n_{i\downarrow} - H/2 \sum_i (n_{i\uparrow} - n_{i\downarrow}) \quad (1)$$

where $\epsilon_i$ is the site energy at the $i$-th site; it takes on the value $\epsilon_A$ or $\epsilon_B$ depending on whether it belongs to an odd ($A$-type) or an even ($B$ type) numbered site respectively. The number operator $n_{i\sigma} = c^\dagger_{i\sigma} c_{i\sigma}$ ($\sigma =\uparrow, \downarrow$) and $n_i = n_{i\uparrow} + n_{i\downarrow}$; $t$ is the hopping integral between the nearest neighbour sites. The phase factor ($e^{i\phi_1}$) appears because of the flux ($=N\phi_1$ in units of basic flux quantum $\phi_0 = hc/2\pi e$) threaded by the ring due to the applied magnetic field. $U$ is the on-site Coulomb interaction. The fourth term represents the Zeeman interaction between the local moments (in units of Bohr magneton) and the applied magnetic field $H$ ($=4\pi\phi_1/N$ for unit lattice spacing). In order to study the conductivity of the present model we calculate the response to a small static electric field in terms of the Drude weight [18, 19]. This amounts to the application of an additional infinitesimally small flux $\phi_2$ (in units of basic flux quantum $\phi_0 = hc/2\pi e$), which modifies the phase factor in the hopping term as well the Zeeman term, and the Hamiltonian looks like :

$$\mathcal{H} = \sum_i \epsilon_i n_i + t \left( e^{i(\phi_1+\phi_2)} \sum_{i\sigma} c^\dagger_{i\sigma} c_{i+1\sigma} + h.c. \right) + U \sum_i n_{i\uparrow} n_{i\downarrow} - H(\phi_2)/2 \sum_i (n_{i\uparrow} - n_{i\downarrow}) \quad (2)$$

On employing the GHFA the Hamiltonian is decoupled into up-spin and down-spin parts with the modified site energies for the up- and the down-spin electrons. The modified site energies are given by:

$$\epsilon'_{i\uparrow} = \epsilon_i + U <n_{i\downarrow}> \ , \ \ \epsilon'_{i\downarrow} = \epsilon_i + U <n_{i\uparrow}> \quad (3)$$

This amounts to a decoupling of the Hamiltonian in to down and up spin parts as follows:

$$\mathcal{H} = \mathcal{H}_\uparrow + \mathcal{H}_\downarrow ,$$

$$\mathcal{H}_\uparrow = \sum_i (\epsilon'_B - H/2)n_{2i\uparrow} + \sum_i (\epsilon'_A - H/2)n_{2i-1\uparrow} + (t\sum_i c^\dagger_{i\uparrow}c_{i+1\uparrow}e^{j\phi} + h.c)$$

$$\mathcal{H}_\downarrow = \sum_i (\epsilon'_B + H/2)n_{2i\downarrow} + \sum_i (\epsilon'_A + H/2)n_{2i-1\downarrow} + (t\sum_i c^\dagger_{i\downarrow}c_{i+1\downarrow}e^{j\phi} + h.c) \quad (4)$$

where $\phi = \phi_1 + \phi_2$. Here $\epsilon'_{B\sigma}$ and $\epsilon'_{A\sigma}$ are determined according to equation (3).

Upon diagonalization (in a self-consistent method) of the decoupled "independent electron" Hamiltonian (4) one can calculate the partition function

$$Z = Tr[e^{-(\beta H - \mu N)}] ,$$

where $\beta = 1/kT$, $k$ being Boltzmann's constant, and $T$, the absolute temperature. $\mu$ is the chemical potential. This leads to direct calculation of several thermodynamic quantities. The Drude weight can be evaluated from

$$D = -\frac{1}{N}\sum_n \left[ \left(\frac{\partial^2 E_{n\uparrow}}{\partial \phi_2^2}\right)_{\phi_2=0} \cdot \frac{1}{e^{\left(\frac{E_{n\uparrow}-\mu}{kT}\right)}+1} + \left(\frac{\partial^2 E_{n\downarrow}}{\partial \phi_2^2}\right)_{\phi_2=0} \cdot \frac{1}{e^{\left(\frac{E_{n\downarrow}-\mu}{kT}\right)}+1} \right] \quad (5)$$

Here $E_{n\sigma}$'s are the single particle energy levels obtained by diagonalizing the decoupled Hamiltonian (4) for the spin species $\sigma$.

The CDW order parameter $C$, the SDW order parameter $S$ and the magnetization $M$ are defined by:

$$C = \frac{1}{N}\sum_i (-1)^i(n_{i\uparrow} + n_{i\downarrow} - 1) \quad (6)$$

$$S = \frac{1}{N}\sum_i (-1)^i(n_{i\uparrow} - n_{i\downarrow}) \quad (7)$$

$$M = \frac{1}{N}\sum_i (n_{i\uparrow} - n_{i\downarrow}) \quad (8)$$

We now discuss the special case of $U = 0$ which is exactly solvable for the infinite system and also gives an insight into the finite size case, for both $U = 0$ and $U > 0$.

In this case ($U=0$) exact diagonalization of the Hamiltonian (4) yields two energy bands for both species of spins. The bands for up and down spin take the following shapes respectively:

$$E(k)_\uparrow = (\epsilon_A + \epsilon_B)/2 - H/2 \pm \frac{1}{2}\sqrt{(\epsilon_B - \epsilon_A)^2 + 8t^2(1 + \cos 2(k+\phi))}$$

$$E(k)_\downarrow = (\epsilon_A + \epsilon_B)/2 + H/2 \pm \frac{1}{2}\sqrt{(\epsilon_B - \epsilon_A)^2 + 8t^2(1 + \cos 2(k + \phi))} \qquad (9)$$

where $k$ is the momentum space index. In absence of the magnetic field $H$ the bands for the two spin species are degenerate. In the special case of $H = 0$ and $\epsilon_A = \epsilon_B$, up and down spins occupy a single band. For $H = 0$ and $\epsilon_A \neq \epsilon_B$, the band gap for each spin species equals $(\epsilon_B - \epsilon_A)$. At half-filling the ground state is obtained by completely filling up the lower bands which leads to the insulating behaviour of the system. Interesting features are expected with the application of a magnetic field in the system. Switching on of $H$ removes the spin degeneracy of the bands. Upon increasing $H$ from zero the upper and lower bands of up-spin go down in energy scale while the corresponding bands of down spin go up. As a result of this there appears a critical value of $H$, $H_c = (\epsilon_B - \epsilon_A)$, where the electrons de-populate the down-spin band and go over to the upper vacant band of up-spin electrons. This process makes the bands partially filled and one expects a metallic behaviour. This aspect must be reflected in the Drude weight ($D$) which is a measure of the dc conductivity. In the next section we discuss the results of the present calculation.

## III. Discussion of Results

We consider the case with $\epsilon_A = 0$ while the parameters $\epsilon_B$, $U$ and $T$ could be varied. Fig.1 shows the Drude weight ($D$) as a function of the magnetic field ($H$) at half-filling with $\epsilon_B = 0$ (and, therefore, $\epsilon_B = \epsilon_A$) and $U = 0$ for different system sizes for low and high temperatures. Let us first consider the low temperature case ($T \approx 0$). Decrease in $D$ with magnetic field $H$ is the characteristic feature of all the three curves corresponding to $N = 54, 102, 154$, which could be anticipated from the energy band equations obtained in the previous section. For $\epsilon_B = 0$, $U=0$ both the species of spins belong to a single band. At half-filling for $H=0$, $D$ is maximum. But with the increase in $H$ the degeneracy of the up and down spins is lifted up. The up-spin band goes down and the down-spin band goes up. So, the uppermost down spin electrons can *"see"* vacant lower energy levels in the up-spin band and emigrate to occupy these levels. Consequently both the bands will be driven away from half-filling, and therefore, $D$ decreases with $H$. The step-like fall and the oscillation in $D$ are finite size effects. Finite system size accounts for the discrete values of momentum index $k$ or, in other words, leads to the discretization of the energy levels. As a result of this, de-excitation (i.e. flipping of the down-spins) occurs at specific values of $H$ which is responsible for step-like fall in $D$. This feature disappears with increasing system size as can be observed from the plot. For a system of infinite size there will be no change in the spectrum with the variation of $\phi_1$ because the spectrum is continuous, there will be merely a change of variable from $k$ to $k'$ ($k'=k+\phi_1$). Whereas, for a finite-size system the spectrum is discrete i.e. $k$ takes on discrete values ($k = 2\pi n/L$, $n = 0, 1, 2, ..., L-1$, where $L$ is the system size). A change in $\phi_1$ will shift an individual energy level to a new value where there was no level existing before. In other words, the spectrum oscillates with $\phi_1$, which is again reflected in $D$. This behaviour also vanishes with increasing system size. On the other hand, there is a smooth fall in $D$ as a function of $H$ for $\epsilon_B = 0, U = 0$ at finite temperature ($kT = 0.1$). The step-like fall and the oscillations, as observed in case of very low temperature, are absent here. This is because at finite temperatures the higher states above the Fermi level are occupied. Therefore, average occupations of up and down spin band change in a continuous fashion with the increase in $H$. Fig.1 suggests that the finite-size effects are almost undetectable even for a system size $N \approx 150$ near zero temperature while such effects are smeared out much more rapidly at high temperatures.

Fig.2 shows the plot of $D$ vs. $H$ for $\epsilon_B = 1$ and $U = 0$ at low and high temperatures. Cases of $N = 54, 102, 154$ are studied at $kT = 0.01$. In this case each of the spin species has two bands and at half-filling the lower bands are completely filled. Therefore, $D$ remains zero from $H = 0$ up to a certain critical value of $H$, say $H_c$. When $H$ reaches the critical value $H_c \approx (\epsilon_B - \epsilon_A)$, the upper vacant band of up-spin gets populated at the cost of depopulation of the down-spin band. This is exactly what we observe in Fig.2. The rise in $D$ with $H$ at higher temperature ($kT = 0.1$), however, is rather gradual. Finite occupation of higher levels at finite temperature is responsible for smoothening of the rise. Also the value of $D$ is higher than that at $T \approx 0$. In Fig.3 we observe the effect of electronic correlation $U$. Introduction of $U$ destroys the CDW ordering induced by the chemical modulation. As a result of this the charge ordering energy gap is reduced, which in turn accounts for the lowering of the critical value of $H$ at $T \approx 0$. At finite temperature, for reasons mentioned earlier, the rise in $D$ is again gradual. Due to thermal excitations across the CDW gap the conductivity is expected to increase with $T$ upto $H \leq H_c(T \approx 0)$ while it would decrease with $T$ for $H > H_c(T \approx 0)$ which already appeared to be metallic ($D \neq 0$) at $T \approx 0$.

Fig.4 shows the variation of $D$ as a function of $U/t$ for different values of $H$ at $T \approx 0$ for $N = 54$. In this plot, a fixed strength of the chemical modulation is considered ($\epsilon_B = 0.5$). It is interesting to note that the rate of change of $D$ is very much dependent on the value of the field $H$. It is really governed by the nature of the ground state as decided by the competition between $\epsilon_B$, $U$ and $H$. This system favours a CDW ground state in absence of correlation and field ($U = 0, H = 0$). In this phase $D$ is zero [17]. For a small value of $H$ (=0.3, well below $H_c$) this CDW ordering is still retained. Now the increase in $U/t$ breaks up the CDW ordering and as a result of this the Drude weight shows a rise upto a certain $U/t = (U/t)_c$ where the SDW instability sets in. In the SDW phase the Drude weight will fall in a characteristic fashion; the peak in $D$ really marks this transition [17]. To verify this conjecture we plot the CDW order parameter ($C$), the SDW order parameter ($S$) and the magnetization ($M$) as functions of $U/t$ in Fig.5(a), for $H = 0.3$ and $\epsilon_B = 0.5$. The CDW order parameter shows a gradual fall with increasing $U/t$ while the magnetization remains zero upto $(U/t)_c$. Beyond this value of $U/t$ the SDW order parameter takes over (Fig.5(a)). Note the "knee" in the plot of $C$ together with a small rise in $M$ associated with the CDW/SDW transition. For a value of $H \approx H_c$, $D \neq 0$ for $U/t = 0$ because of depopulation of down-spin band as mentioned earlier. Hence, the rate of increase of $D$ with $U/t$ is of much appreciable magnitude (the case of $H = 0.5$ in Fig.4). This is because, at this value of $H$, the overlap between the lower band of the down spins and the upper band of the up spins increases. Therefore, with the increase in $U/t$, the double occupancies are broken and simultaneous depopulation of the down-spin band takes place. This enhances the conductivity. This is reflected in Fig.4 ($H = 0.5$) as well as in the corresponding plots of $C$ and $M$ as functions of $U/t$ in Fig.5(b). It can be seen from Fig.4 that $D$ saturates to a high value over a small range of $U/t$. This region on the $U/t$ axis essentially corresponds to a magnetically polarized phase as revealed in Fig.5(b) which shows comparable values of $C$ and $M$ in this regime.

Much higher values of $U/t$ correspond to an SDW phase where the Drude weight falls sharply ($H = 0.5$ in Fig.4) with a fast rise in $S$ (Fig.5(b)). However, for very large values of $H$ ($> H_c$), the depopulation of the down-spin band is nearly complete. So very small changes in $C$ and $M$ (as functions of $U/t$) are expected in the CDW region and this is really observed in Fig.5(c). Consequently the Drude weight shows a very small increase in its value as plotted against $U/t$ in Fig.4 (the case of $H = 0.8$). Here, in Fig.5(c), we can identify two values of $U/t$, say $(U/t)_{c1}$ and $(U/t)_{c2}$ between which $M$ clearly dominates over $C$ indicating a phase of strong magnetic polarization. In the corresponding region in Fig.4 ($H = 0.8$) we find $D$ saturating to its maximum. Beyond $(U/t)_{c2}$, however, we find usual drop in $D$ (Fig.4) together with a sharp rise in $S$ in the SDW sector. We note that the changes in the order parameters as well as the Drude weight become sharper with larger values of the field.

## IV. Conclusion.

Summarizing, we have studied the conductivity of finite-sized Hubbard ring threaded by a magnetic flux in presence of an alternating modulation in the site potentials. Systems at half-filling are considered at low and high temperatures. We have considered system sizes $N \approx 100$ and found that the finite size effects in Drude weight disappear for systems consisting of nearly 150 sites at low temperatures. The competition between the correlation, the magnetic field and the chemical modulation tune the conductivity in an interesting fashion. In absence of correlation and modulation in site potentials there is a slow fall in the Drude weight as a function of the external field. The oscillations and step-like fall in the Drude weight quickly disappears with increasing system size. Upon introduction of the chemical modulation there arises a critical value of the magnetic field across which there appears a sharp rise in the conductivity. This accounts for a transition from an insulating phase to a conducting one. However, the introduction of the Hubbard correlation tends to destroy the charge ordering effect induced by the chemical modulation and, consequently, reduces the value of the critical field. This accounts for an enhancement of conductivity as well. The critical field, however, is found to be insensitive to the system size. The competition between the three factors are visible from the variation of the SDW and CDW order parameters and the magnetization of the system. Transition from a CDW to an SDW phase (at low field) or that from a CDW to a magnetically polarized one to an SDW (at high field) are found to have notable effects on the conductivity of the system. At finite temperatures the nature of variation of the conductivity with the external field is much similar to that observed for $T \approx 0$ but the rise or fall in $D$ is much more gradual. Also the finite size effects are not visible even for a system size of $N \approx 50$ at higher temperatures. It seems from the present study that the interplay of correlation and chemical modulation brings in observable features in the magnetoconductance in a system. Effects of variation in band filling and the role of diagonal/off-diagonal disorder could be interesting areas of further study of this problem.

# Figure captions:

**Fig.1**: Plot of the Drude weight $D$ as a function of magnetic field $H$ for three different system sizes: $N = 54, 102$ and $154$ at $kT = 0.01$, and for two different system sizes $N = 54$ and $102$ at $kT=0.1$ for $\epsilon_B = 0.0, U = 0.0$. The scale of energy is $t = 1.0$. $\epsilon_A$ is chosen to be zero.

**Fig.2**: Plot of $D$ versus $H$ for $N = 54, 102$ and $154$ at $kT = 0.01$, and for $N=54$ and $102$ at $kT=0.1$ for $\epsilon_B = 1.0, U = 0.0$. The scale of energy is $t = 1.0$. $\epsilon_A$ is chosen to be zero.

**Fig.3**: Plot of $D$ versus $H$ for $N = 54, 102$ and $154$ at $kT=0.01$, and for $N=54$ and $102$ at $kT=0.1$ for $\epsilon_B = 1.0, U = 1.0$. The scale of energy is $t = 1.0$. $\epsilon_A$ is chosen to be zero.

**Fig.4**: Variation of $D$ as a function of $U/t$ at a fixed value of chemical modulation $\epsilon_B = 0.5$ for $H = 0.3, 0.5$ and $0.8$ at $kT=0.01$. The ring consists of 54 sites. $\epsilon_A$ is chosen to be zero.

**Fig.5**: Plots of CDW order parameter $C$, magnetization $M$ and SDW order parameter $S$ against $U/t$ for (a) $H = 0.3$, (b) $H = 0.5$ and (c) $H = 0.8$ at $kT = 0.01$. The ring consists of 54 sites. $\epsilon_B = 0.5$ and $\epsilon_A = 0$.

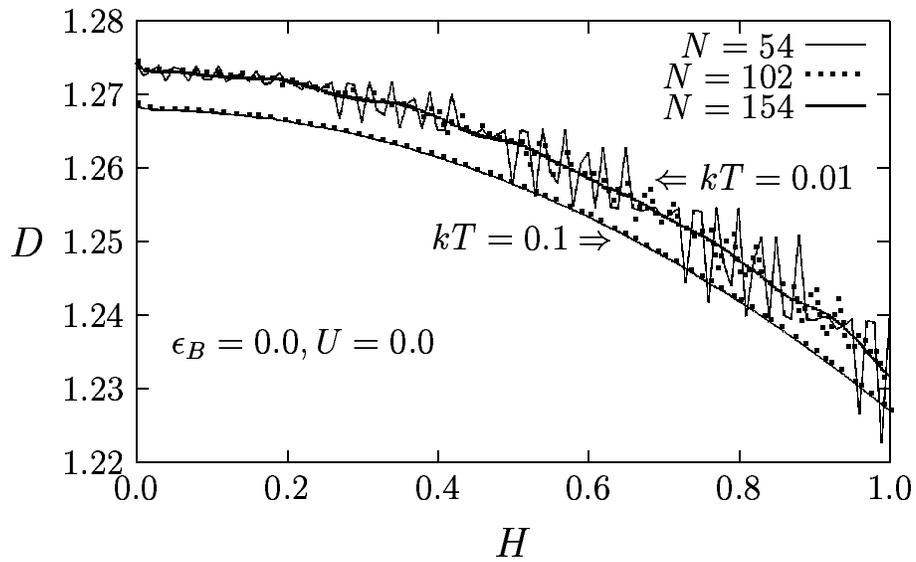

Fig.1

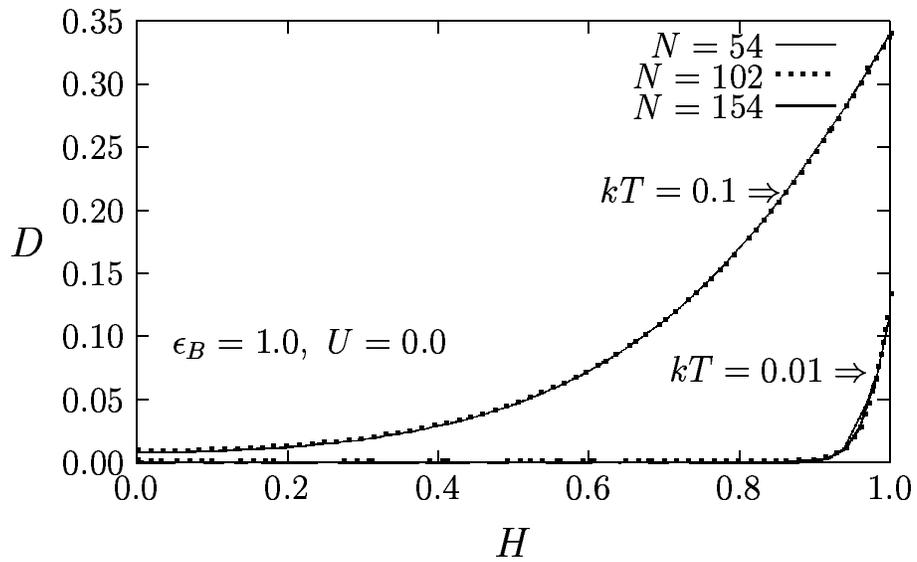

Fig. 2

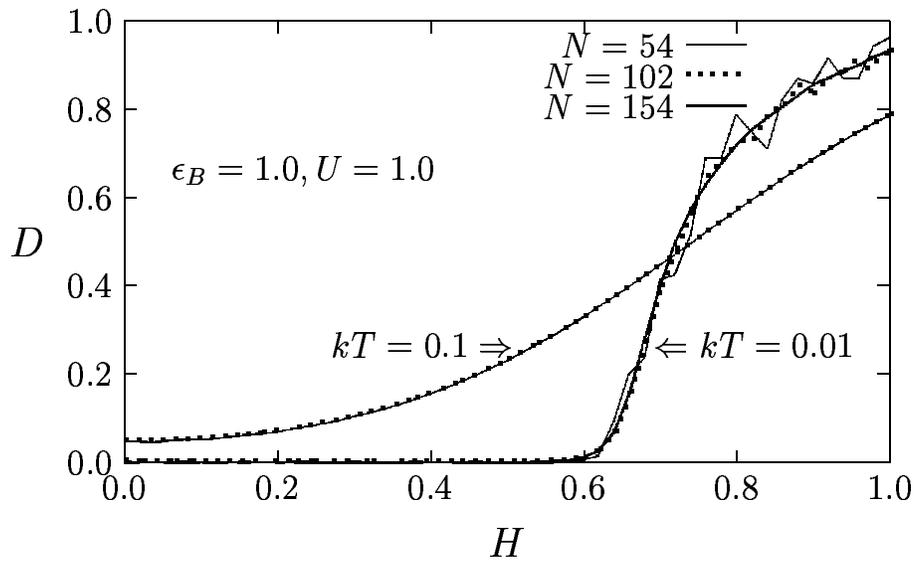

Fig. 3

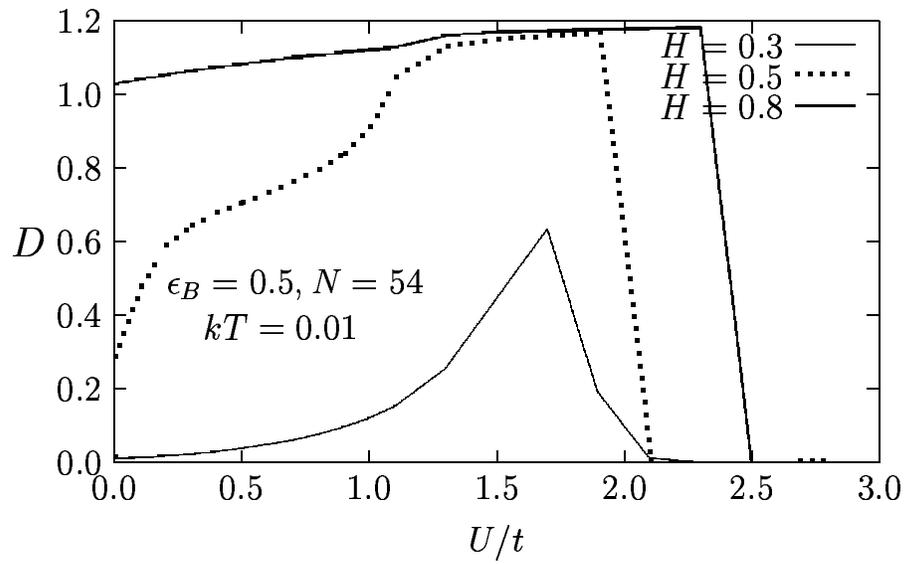

Fig. 4

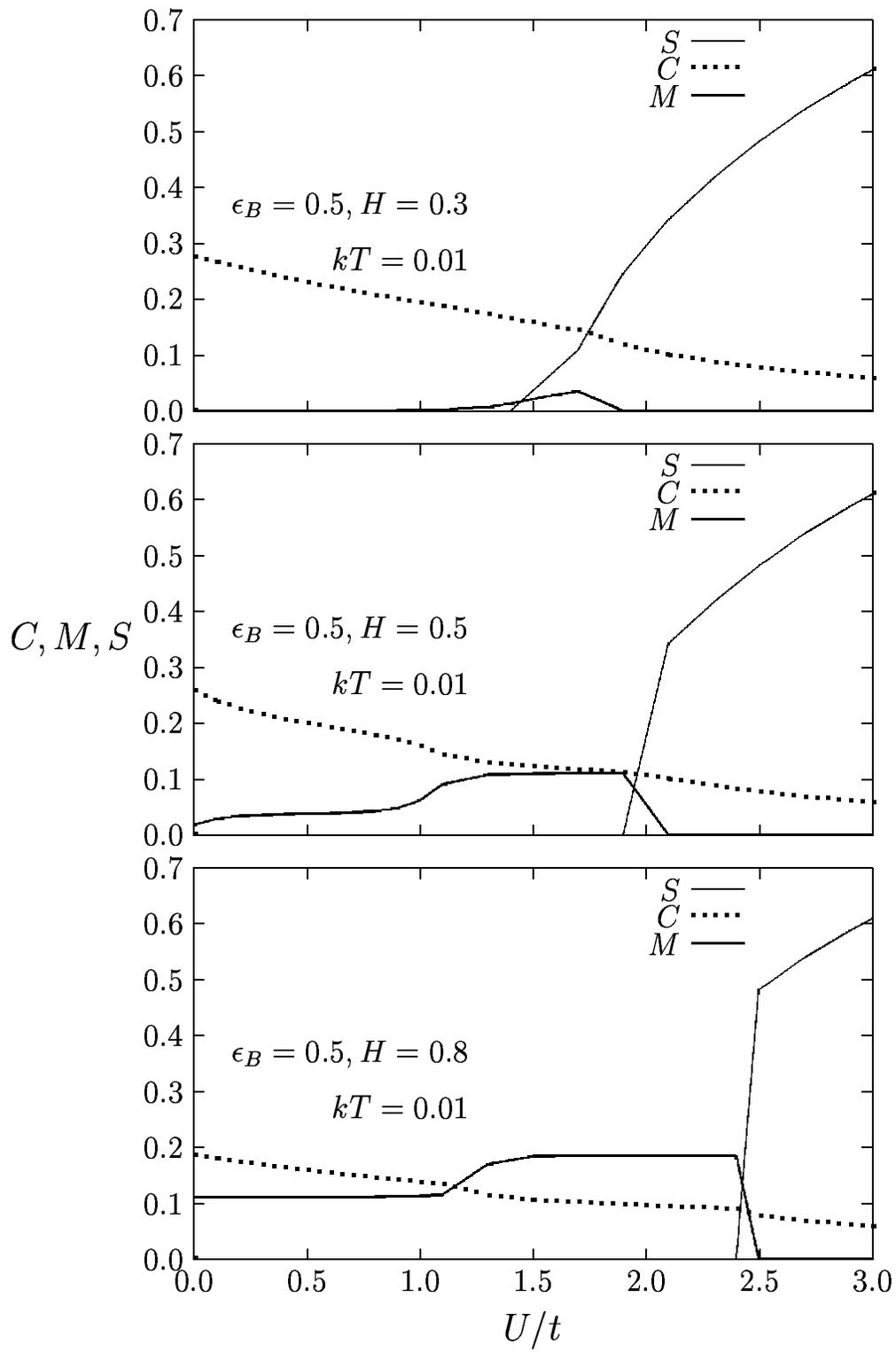

Fig. 5